\documentclass[pdflatex,sn-mathphys-num]{sn-jnl}

\usepackage{graphicx}%
\usepackage{multirow}%
\usepackage{amsmath,amssymb,amsfonts}%
\usepackage{amsthm}%
\usepackage{mathrsfs}%
\usepackage[title]{appendix}%
\usepackage{xcolor}%
\usepackage{textcomp}%
\usepackage{manyfoot}%
\usepackage{listings}%
\usepackage{bm}
\newcommand{\csch}{\mathrm{csch}}
\begin{document}

\title[Stress Analysis of a Square Elastic Body Under Biaxial Loading Using Airy Stress Functions]{Stress Analysis of a Square Elastic Body Under Biaxial Loading Using Airy Stress Functions}

\author[1]{\fnm{Ryu} \sur{Suzuki}}
\email{s243124y@st.go.tuat.ac.jp}

\author[2]{\fnm{Shintaro} \sur{Hokada}}
\email{hokada@st.go.tuat.ac.jp}

\author*[1]{\fnm{Satoshi} \sur{Takada}}
\email{takada@go.tuat.ac.jp}

\affil[1]{\orgdiv{Department of Mechanical Systems Engineering}, 
    \orgname{Tokyo University of Agriculture and Technology}, 
    \orgaddress{
    \street{2-24-16 Naka-cho}, 
    \city{Koganei}, 
    \postcode{184-8588}, 
    \state{Tokyo}, 
    \country{Japan}}}
\affil[2]{\orgdiv{Department of Industrial Technology and Innovation}, 
    \orgname{Tokyo University of Agriculture and Technology}, 
    \orgaddress{
    \street{2-24-16 Naka-cho}, 
    \city{Koganei}, 
    \postcode{184-8588}, 
    \state{Tokyo}, 
    \country{Japan}}}

\abstract{
This study presents an analytical investigation of stress distributions in square-shaped elastic bodies subjected to concentrated compressive loads under uniaxial and biaxial conditions. 
By employing the Airy stress function method, we derive closed-form solutions that satisfy the governing biharmonic equation and the prescribed boundary conditions along the edges of the square domain. 
The stress components are expressed as series expansions, with coefficients determined to enforce boundary constraints.
In the uniaxial compression case, the resulting stress fields exhibit strong agreement with photoelastic fringe patterns previously observed in experimental studies. 
For biaxial loading, the solution represents a superposition of two orthogonal compression scenarios, producing spatial variations in the principal stress difference depending on the location within the domain.
}

\keywords{Biaxial loading, Elasticity, Airy stress function}

\maketitle

\section{Introduction}\label{sec:introduction}
The field of two-dimensional elasticity plays a central role in theoretical solid mechanics and continues to provide valuable tools for understanding stress and strain distributions in structural elements. 
Despite the rise of powerful numerical methods, exact and semi-analytical solutions in elasticity remain essential for benchmarking, theoretical validation, and educational purposes.
Moreover, the development of analytical solutions often reveals underlying mechanical principles that may not be immediately visible through numerical approximations alone.

Classical problems—such as stress analysis in circular or elliptical domains—have long served as canonical benchmarks for validating computational tools and understanding physical behavior~\cite{Timoshenko70, Coker57, Fung01, Lu18, Guerrero-Miguel19, Ramesh22, Shins23}. 
A well-known example involves a circular disk subjected to diametrically opposed point loads, for which closed-form solutions can be derived using Airy stress functions in polar coordinates~\cite{Timoshenko70, Coker57}. 
These results are not only of theoretical importance but also serve as reference standards for experimental techniques, such as photoelasticity~\cite{Frocht31, Coker57, Ramesh20}, and are widely used in engineering applications including fracture mechanics and contact problems.

The dynamic counterpart of such problems has also been studied extensively~\cite{Jingu85_2D, Jingu85_3D, Kessler91, Sato24_2D, Sato24_3D}, enhancing our understanding of wave propagation in elastic solids. 
These analyses are particularly relevant in seismology, material characterization, and non-destructive evaluation.

However, analytical solutions for domains with non-circular boundaries, especially polygonal geometries such as squares and rectangles, remain relatively scarce. 
The absence of radial symmetry, combined with the complexity of applying realistic boundary conditions, often precludes the use of separable coordinate systems. 
Nevertheless, these geometries frequently appear in practical scenarios—e.g., square specimens are commonly employed in indirect tensile strength evaluations like the Brazilian test~\cite{Zhao19, Manjit21, Saito52, Shimada54}.

Although finite element methods (FEM) are routinely applied to such problems, they are inherently approximate and depend on mesh quality and boundary representation. 
Thus, closed-form or semi-analytical solutions, when obtainable, remain valuable both as benchmarks and as tools for gaining physical insight.

Previous studies have extended the use of Airy stress functions to non-standard geometries, including octagonal~\cite{Manjit24}, rhomboidal~\cite{Benedetti19, Segura21}, and perforated circular domains~\cite{Okamura25}, demonstrating the method's flexibility. 
Yet, rigorous analytical treatments of square-shaped bodies under complex boundary conditions and loading configurations remain limited.

Beyond its theoretical significance, the analytical framework developed here is directly relevant to experimental configurations that impose biaxial compression on square or rectangular specimens.
Such loading conditions arise in various mechanical testing setups, for instance in multi-axial loading rigs used to investigate plasticity and fracture in sheet metals and other engineering materials~\cite{Kuwabara07}.
In these experiments, accurately capturing the stress state within the specimen is essential for interpreting measured strains, identifying yield criteria, and calibrating constitutive models.
By providing closed-form or semi-analytical solutions under both uniaxial and biaxial compressive loadings, the present work offers benchmark data that can complement and validate such experimental investigations.

In this context, this study addresses an important gap by providing an analytical framework for evaluating stress distributions in square elastic domains subjected to concentrated and distributed compressive loads under both uniaxial and biaxial conditions. 
The analysis is performed within the context of two-dimensional linear elasticity, which, despite its simplifying assumptions, captures essential stress behavior in many thin-plate structures and serves as a stepping stone to more complex three-dimensional models. 
The choice of two-dimensional elasticity is justified by its mathematical tractability and its effectiveness in modeling plane stress and plane strain scenarios commonly encountered in engineering.

The Airy stress function is employed in this study due to its ability to reduce the governing partial differential equations of elasticity to a single biharmonic equation, which facilitates the construction of general solutions while inherently satisfying equilibrium conditions~\cite{Timoshenko70, Fung01}. 
This method allows us to explore the effects of different boundary conditions—such as simply supported and traction-free edges—and loading schemes, including both point and uniform compressive loads.
Unlike many previous works that focus on numerical or approximate solutions, our approach seeks to develop exact or semi-analytical expressions by satisfying the biharmonic condition and enforcing boundary compatibility via Fourier series expansions and polynomial approximations.

While the subject of stress analysis in two-dimensional materials is not entirely new, the novelty of this work lies in the construction of generalized Airy stress functions tailored to square geometries with realistic boundary conditions, which are not readily addressed by classical methods. 
In particular, we systematically derive stress fields under various loading scenarios and analyze their convergence behavior and potential use in validating computational models.
Ultimately, this study aims to contribute both to the theoretical development of elasticity and to its practical applications, including structural optimization, failure prediction, and experimental validation. 
The results presented here may also serve as a reference for future extensions to rectangular, anisotropic, or layered domains, as well as for exploring dynamic and three-dimensional elasticity problems.

\section{Setup}\label{sec:setup}
In this section, we present the formulation of the problem and describe the model geometry and boundary conditions.
\begin{figure}[htbp]
    \centering
    \includegraphics[width=0.75\linewidth]{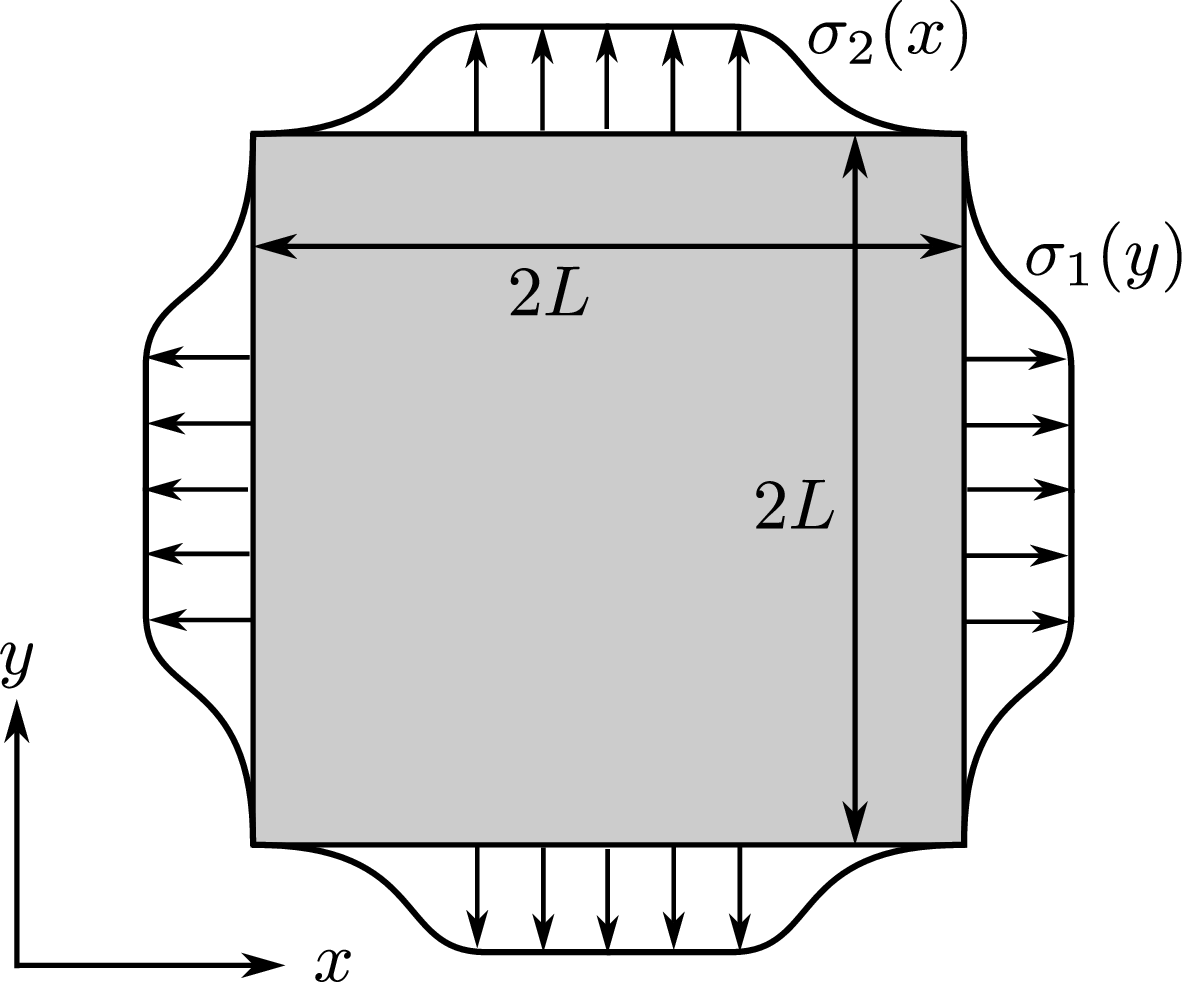}
    \vspace{1.5em}
    \caption{Schematic of the system.
    The external stresses $\sigma_1(y)$ and $\sigma_2(x)$ act on the outer edges of an elastic body with a square section whose linear size is $2L$.}
    \label{fig:setup}
\end{figure}

As illustrated in Fig.~\ref{fig:setup}, we consider a homogeneous, isotropic, and linearly elastic body with a square cross-section of side length $2L$. 
For convenience and symmetry, the Cartesian coordinate origin is placed at the center of the square domain.
External normal stresses, denoted by $\sigma_1(y)$ and $\sigma_2(x)$, are applied to the lateral edges at $x=\pm L$ and the top and bottom edges at $y=\pm L$, respectively. 
These stress distributions may represent concentrated, distributed, or mixed loading conditions. The boundary conditions on all four sides of the square are assumed to be either simply supported or traction-free in the tangential direction.

The boundary conditions on the square edges are expressed as follows:
\begin{subequations}\label{eq:BC}
\begin{align}
    \sigma_{xx}(\pm L, y) &= \sigma_1(y), \quad
    \sigma_{xy}(\pm L, y) = 0, \\
    \sigma_{yy}(x, \pm L) &= \sigma_2(x), \quad
    \sigma_{xy}(x, \pm L) = 0,
\end{align}
\end{subequations}
where $\sigma_1(y)$ and $\sigma_2(x)$ represent the externally applied normal stress distributions along the vertical and horizontal edges, respectively. 
The shear stress components $\sigma_{xy}$ vanish on all boundaries, reflecting either frictionless contact or traction-free conditions.

To facilitate analytical treatment, the external stress distributions are expanded in terms of cosine Fourier series:
\begin{subequations}\label{eq:sigma1_sigma2}
\begin{align}
    \sigma_1(y)
    &= \frac{\widetilde{\sigma}_1^{(0)}}{2}
    + \sum_{n=1}^\infty \widetilde{\sigma}_1^{(n)} 
    \cos\left(\frac{n\pi}{L}y\right),\\
    \sigma_2(x)
    &= \frac{\widetilde{\sigma}_2^{(0)}}{2}
    + \sum_{n=1}^\infty \widetilde{\sigma}_2^{(n)} 
    \cos\left(\frac{n\pi}{L}x\right),
\end{align}
\end{subequations}
where $\widetilde{\sigma}_1^{(n)}$ and $\widetilde{\sigma}_2^{(n)}$ are the Fourier coefficients representing the spectral content of the stress distributions along each edge. 
These expansions allow for systematic satisfaction of boundary conditions and form the basis for the construction of compatible Airy stress functions in subsequent analysis.

In the following sections, we derive the internal stress fields within the square domain that satisfy both the governing biharmonic equation and the boundary conditions given in Eqs.~\eqref{eq:BC}--\eqref{eq:sigma1_sigma2}. 
The Airy stress function method provides a powerful analytical framework for this purpose, reducing the problem to the construction of biharmonic functions whose second derivatives yield physically meaningful stress components.

\section{Airy stress function}\label{sec:Airy}
Let us consider the force balance equation:
\begin{equation}
    \bm\nabla \cdot \overleftrightarrow{\sigma} = \bm{0},
\end{equation}
where $\overleftrightarrow{\sigma}$ is the stress tensor of the system.
In two-dimensional linear elasticity, the stress components can be conveniently expressed in terms of the Airy stress function $\phi(x,y)$, which automatically satisfies the equilibrium equations in the absence of body forces~\cite{Timoshenko70, Coker57}. 
Specifically, the stress components are related to $\phi(x,y)$ as follows:
\begin{equation}
    \sigma_{xx}(x, y)
    = \frac{\partial^2\phi(x, y)}{\partial y^2},\quad
    \sigma_{yy}(x, y)
    = \frac{\partial^2\phi(x, y)}{\partial x^2},\quad
    \sigma_{xy}(x, y)
    = -\frac{\partial^2\phi(x, y)}{\partial x \partial y}.
    \label{eq:sigma_phi}
\end{equation}

In order for the stress field to be physically admissible (i.e., compatible with strain and displacement fields), the Airy stress function must satisfy the biharmonic equation:
\begin{equation}
    \nabla^4\phi
    = \nabla^2 \left(\nabla^2\phi\right)
    = 0.
    \label{eq:biharmonic}
\end{equation}
Although the general solution to Eq.~\eqref{eq:biharmonic} is known~\cite{Timoshenko70}, it suffices for our purposes to consider a solution in the following separable form:
\begin{align}
    \phi(x, y)
    &= \frac{E}{2}x^2+\frac{F}{2}y^2
    + \sum_{n=1}^\infty 
    \cos\left(\frac{n\pi}{L}x\right)
    \left[A_n\cosh\left(\frac{n\pi}{L}y\right)
    +B_n \frac{n\pi}{L}y\sinh\left(\frac{n\pi}{L}y\right)\right]\nonumber\\
    &\hspace{1em}
    + \sum_{n=1}^\infty 
    \left[C_n\cosh\left(\frac{n\pi}{L}x\right)
    +D_n \frac{n\pi}{L}x \sinh\left(\frac{n\pi}{L}x\right)\right]
    \cos\left(\frac{n\pi}{L}y\right),
    \label{eq:Airy}
\end{align}
where $A_n$, $B_n$, $C_n$, $D_n$, $E$, and $F$ are constants to be determined from the boundary conditions specified in Eq.~\eqref{eq:BC}. 
See Appendix \ref{sec:biharmonic} for the detailed derivation.
This representation reflects the symmetries of the problem and facilitates separation of variables.

By differentiating Eq.~\eqref{eq:Airy}, we obtain the corresponding stress components:
\begin{subequations}\label{eq:stress}
\begin{align}
    \sigma_{xx}(x,y)
    &= F 
    - \sum_{n=1}^\infty \left(\frac{n\pi}{L}\right)^2
    \left[C_n\cosh\left(\frac{n\pi}{L}x\right)
    +D_n \frac{n\pi}{L}x \sinh\left(\frac{n\pi}{L}x\right)\right]
    \cos\left(\frac{n\pi}{L}y\right)\nonumber\\
    &\hspace{1em}
    + \sum_{n=1}^\infty 
    \left(\frac{n\pi}{L}\right)^2
    \cos\left(\frac{n\pi}{L}x\right)\nonumber\\
    &\hspace{2em}\times
    \left\{A_n\cosh\left(\frac{n\pi}{L}y\right)
    + B_n
    \left[\frac{n\pi}{L} y 
    \sinh\left(\frac{n\pi}{L}y\right)
    +2\cosh \left(\frac{n\pi}{L}y\right)\right]\right\},\\
    \sigma_{yy}(x, y)
    &= E
    - \sum_{n=1}^\infty 
    \left(\frac{n\pi}{L}\right)^2
    \cos\left(\frac{n\pi}{L}x\right)
    \left[A_n\cosh\left(\frac{n\pi}{L}y\right)
    +B_n \frac{n\pi}{L}y\sinh\left(\frac{n\pi}{L}y\right)\right]\nonumber\\
    &\hspace{1em}
    + \sum_{n=1}^\infty 
    \left(\frac{n\pi}{L}\right)^2
    \cos\left(\frac{n\pi}{L}y\right)\nonumber\\
    &\hspace{2em}\times
    \left\{C_n\cosh\left(\frac{n\pi}{L}x\right)
    +D_n \left[\frac{n\pi}{L}x \sinh\left(\frac{n\pi}{L}x\right)
    + 2\cosh\left(\frac{n\pi}{L}x\right)\right]\right\},\\
    \sigma_{xy}(x, y)
    &= \sum_{n=1}^\infty
    \left(\frac{n\pi}{L}\right)^2
    \sin\left(\frac{n\pi}{L}x\right)\nonumber\\
    &\hspace{2em}\times
    \left\{A_n\sinh\left(\frac{n\pi}{L}y\right)
    +B_n \left[\sinh\left(\frac{n\pi}{L}y\right)
    + \frac{n\pi}{L}y\cosh\left(\frac{n\pi}{L}y\right)\right]\right\}\nonumber\\
    &\hspace{1em}
    + \sum_{n=1}^\infty 
    \left(\frac{n\pi}{L}\right)^2
    \sin\left(\frac{n\pi}{L}y\right)\nonumber\\
    &\hspace{2em}\times
    \left\{C_n\sinh\left(\frac{n\pi}{L}x\right)
    +D_n \left[\sinh\left(\frac{n\pi}{L}x\right)
    + \frac{n\pi}{L}x \cosh\left(\frac{n\pi}{L}x\right)\right]\right\}.
\end{align}
\end{subequations}

The shear stress component $\sigma_{xy}$ must vanish along the outer edges of the square, i.e., at $x=\pm L$ and $y=\pm L$, in accordance with the boundary conditions. 
By applying these conditions, the coefficients $B_n$ and $D_n$ can be expressed in terms of $A_n$ and $C_n$, respectively:
\begin{subequations}
\begin{align}
    B_n
    &= -\frac{1}{1 + n\pi \coth(n\pi)} A_n, \\
    D_n
    &= -\frac{1}{1 + n\pi \coth(n\pi)} C_n.
\end{align}
\end{subequations}
These relations ensure that the shear stress components are identically zero on all edges, satisfying the tangential traction-free conditions.


Let us now apply the boundary conditions~\eqref{eq:BC} for the normal stress components $\sigma_{xx}$ and $\sigma_{yy}$.
To this end, we expand the functions $\cosh(x)$ and $x\sinh(x)$ in Fourier cosine series as follows:
\begin{subequations}\label{eq:Fourier}
\begin{align}
    \cosh\left(\frac{n\pi}{L}x\right)
    &= \frac{c_{n,0}}{2}
    + \sum_{m=1}^\infty 
    c_{n,m} \cos\left(\frac{m\pi}{L}x\right),\\
    \frac{n\pi}{L}x \sinh\left(\frac{n\pi}{L}x\right)
    &= \frac{s_{n,0}}{2}
    + \sum_{m=1}^\infty 
    s_{n,m} \cos\left(\frac{m\pi}{L}x\right).
\end{align}
\end{subequations}
Here, the Fourier coefficients $c_{n,m}$ and $s_{n,m}$ are explicitly given by
\begin{subequations}
\begin{align}
    c_{n,m} 
    &= \frac{(-1)^m}{\pi}\frac{2n \sinh(n\pi)}{m^2+n^2},\\
    s_{n,m}
    &= \frac{(-1)^m}{\pi}
    \frac{2n \left[\sinh(n\pi) (m^2-n^2) 
    +n \pi \cosh(n\pi) (m^2+n^2) \right]}{(m^2+n^2)^2}.
\end{align}
\end{subequations}
By substituting Eqs.~\eqref{eq:Fourier} into the expressions for the stress components given in Eqs.~\eqref{eq:stress}, we obtain the following representations at the boundaries:
\begin{subequations}\label{eq:sigma_L}
\begin{align}
    \sigma_{xx}(L,y)
    &= F
    - \left(\frac{\pi}{L}\right)^2
    \sum_{n=1}^\infty 
    \left(\sum_{m=1}^\infty 
    P_{n,m}A_m + Q_n C_n\right)
    \cos\left(\frac{n\pi}{L}y\right),\\
    \sigma_{yy}(x, L)
    &= E
    - \left(\frac{\pi}{L}\right)^2
    \sum_{n=1}^\infty 
    \left(\sum_{m=1}^\infty P_{n,m}C_m + Q_nA_n\right)\cos\left(\frac{n\pi}{L}x\right).
\end{align}
\end{subequations}
In the above expressions, $P_{n,m}$ and $Q_n$ are defined as
\begin{subequations}
\begin{align}
    P_{n,m}
    &\equiv 
    (-1)^m m^2\frac{s_{m,n}+ c_{m,n}[1-m\pi \coth(m\pi)]}{1+m\pi \coth(m\pi)},\\
    Q_n
    &\equiv 
    n^2\frac{\cosh(n\pi)+n\pi \csch(n\pi)}
    {1+n\pi \coth(n\pi)}.
\end{align}
\end{subequations}
The stress components in Eqs.~\eqref{eq:sigma_L} must match the boundary conditions specified in Eq.~\eqref{eq:BC}, along with the Fourier-expanded forms of the boundary stresses in Eqs.~\eqref{eq:sigma1_sigma2}.
This requirement yields the following set of equations:
\begin{subequations}\label{eq:sim_eqs}
\begin{align}
    F
    = \frac{\widetilde{\sigma}_1^{(0)}}{2},\quad
    E
    &= \frac{\widetilde{\sigma}_2^{(0)}}{2},\\
    \sum_{m=1}^\infty 
    P_{n,m}A_m + Q_n C_n
    &= - \left(\frac{L}{\pi}\right)^2\widetilde{\sigma}_1^{(n)},
    \label{eq:sim_eqs2}\\
    \sum_{m=1}^\infty P_{n,m}C_m + Q_nA_n
    &= - \left(\frac{L}{\pi}\right)^2\widetilde{\sigma}_2^{(n)}.
    \label{eq:sim_eqs3}
\end{align}
\end{subequations}
Equations~\eqref{eq:sim_eqs2} and~\eqref{eq:sim_eqs3} can thus be viewed as a coupled system of linear equations for the unknown coefficients $A_n$ and $C_n$.
Once a solution to this system is found, the full stress field can be reconstructed from Eqs.~\eqref{eq:stress}.

In practice, however, the coefficients $P_{n,m}$ and $Q_n$ diverge for large $n$ and $m$, i.e., $n\gg 1$ or $m\gg 1$.
To suppress this divergence and improve numerical stability, we introduce rescaled variables $\mathcal{A}_n$ and $\mathcal{C}_n$ defined as
\begin{equation}
    \mathcal{A}_n 
    \equiv n^2 \mathrm{e}^{n\pi}A_n,\quad
    \mathcal{C}_n 
    \equiv n^2 \mathrm{e}^{n\pi}C_n.
    \label{eq:cal_A_C}
\end{equation}
Correspondingly, we define modified coefficients $\mathcal{P}_{n,m}$ and $\mathcal{Q}_n$ as
\begin{equation}
    \mathcal{P}_{n,m}
    \equiv \frac{1}{m^2}\mathrm{e}^{-m\pi}
    F_{n,m},\quad
    \mathcal{Q}_n
    \equiv \frac{1}{n^2}\mathrm{e}^{-n\pi}
    Q_n.
\end{equation}
With these new variables, the system of equations~\eqref{eq:sim_eqs} can be rewritten in a numerically stable form:
\begin{align}
    \sum_{m=1}^\infty 
    \mathcal{P}_{n,m}\mathcal{A}_m 
    + \mathcal{Q}_n \mathcal{C}_n
    &= - \left(\frac{L}{\pi}\right)^2\widetilde{\sigma}_1^{(n)},\\
    \sum_{m=1}^\infty \mathcal{P}_{n,m}\mathcal{C}_m 
    + \mathcal{Q}_n\mathcal{A}_n
    &= - \left(\frac{L}{\pi}\right)^2\widetilde{\sigma}_2^{(n)}.
\end{align}
To make the system finite and solvable numerically, we introduce a truncation at $N_\mathrm{max}$ for the indices $n$ and $m$.

Let us now define the vector $\bm{X}$, which collects the unknown coefficients, and the vector $\bm{\Sigma}$, which contains the Fourier components of the boundary stresses:
\begin{align}
    \bm{X}&\equiv 
    \begin{pmatrix}
        \mathcal{A}_1, \mathcal{A}_2,\cdots
        \mathcal{A}_{N_\mathrm{max}},
        \mathcal{C}_1, \mathcal{C}_2,\cdots
        \mathcal{C}_{N_\mathrm{max}}
    \end{pmatrix}^\mathrm{T},\\
    \bm{\Sigma}&\equiv 
    \begin{pmatrix}
        \widetilde{\sigma}_1^{(1)},
        \widetilde{\sigma}_1^{(2)},\cdots
        \widetilde{\sigma}_1^{(N_\mathrm{max})},
        \widetilde{\sigma}_2^{(1)},
        \widetilde{\sigma}_2^{(2)},\cdots
        \widetilde{\sigma}_2^{(N_\mathrm{max})}
    \end{pmatrix}^\mathrm{T}.
\end{align}
Next, we introduce the matrix $\overleftrightarrow{\mathcal{M}}$
 , which consists of block matrices $\overleftrightarrow{\mathcal{P}}$ and $\overleftrightarrow{\mathcal{Q}}$ defined by
\begin{equation}
    \overleftrightarrow{\mathcal{M}}\equiv
    \begin{pmatrix}
        \overleftrightarrow{\mathcal{P}} & \overleftrightarrow{\mathcal{Q}}\\
        \overleftrightarrow{\mathcal{Q}} & \overleftrightarrow{\mathcal{P}}
    \end{pmatrix},
\end{equation}
with 
\begin{equation}
    \overleftrightarrow{\mathcal{P}}\equiv
    \begin{pmatrix}
        \mathcal{P}_{1,1} & \mathcal{P}_{1,2} & \mathcal{P}_{1,3} & \cdots \\
        \mathcal{P}_{2,1} & \mathcal{P}_{2,2} & \mathcal{P}_{2,3} & \cdots \\
        \mathcal{P}_{3,1} & \mathcal{P}_{3,2} & \mathcal{P}_{3,3} & \cdots \\
        \vdots & \vdots & \vdots & \ddots 
    \end{pmatrix},\quad
    \overleftrightarrow{\mathcal{Q}}\equiv
    \begin{pmatrix}
        \mathcal{Q}_1 & 0 & 0 & \cdots \\
        0 & \mathcal{Q}_2 & 0 & \cdots \\
        0 & 0 & \mathcal{Q}_3 & \cdots \\
        \vdots & \vdots & \vdots & \ddots 
    \end{pmatrix}.
\end{equation}
Using this notation, the full system of linear equations can be expressed in compact matrix form as
\begin{equation}
    \bm{X}
    = -\left(\frac{L}{\pi}\right)^2
    \overleftrightarrow{\mathcal{M}}^{-1}
    \bm{\Sigma}.
    \label{eq:X_solution}
\end{equation}
Once the coefficients $\mathcal{A}_n$ and $\mathcal{C}_n$ are determined from Eq.\eqref{eq:X_solution}, we can readily compute the stress components throughout the domain by substituting into Eqs.\eqref{eq:stress}.

This completes the analytical formulation of the stress field in a square elastic body under arbitrary symmetric biaxial loading.
In the following section, we present the results obtained using this framework for several representative loading conditions, highlighting the spatial distributions and convergence behavior of the computed stress components.

\section{Results and discussion}\label{sec:results}
In this section, we present the stress distributions obtained using the analytical method described in the previous section. 
Three representative loading conditions are examined: uniform biaxial tension, uniaxial compression under concentrated loading, and biaxial compression. 
For each case, we evaluate the solution behavior and discuss the physical implications of the resulting stress fields.

\subsection{Validation: Uniform biaxial tension}\label{sec:uniform}
We first consider the simplest case of uniform biaxial tension given by
\begin{equation}
    \sigma_1(y)=\sigma_1,\quad
    \sigma_2(x)=\sigma_2.
\end{equation}
The Fourier coefficients of the boundary conditions are obtained as
\begin{equation}
    \widetilde{\sigma}_1^{(0)}
    = 2\sigma_1,\quad
    \widetilde{\sigma}_2^{(0)} 
    = 2\sigma_2,\quad
    \widetilde{\sigma}_1^{(n)}
    = \widetilde{\sigma}_2^{(n)}
    = 0\ (n\ge 1).
\end{equation}

Solving Eqs.~\eqref{eq:sim_eqs}, we find the following solutions:
\begin{equation}
    E = \sigma_2,\quad
    F = \sigma_1,\quad
    A_1 = A_2 = \cdots = C_1 = C_2 = \cdots = 0.
\end{equation}
Thus, the Airy stress function reduces to a quadratic form, and the stress components are spatially uniform:
\begin{subequations}
\begin{align}
    \phi(x, y)
    &= \frac{\sigma_2}{2}x^2+\frac{\sigma_1}{2}y^2,\\
    \sigma_{xx}(x,y)
    &= \sigma_1,\quad
    \sigma_{yy}(x, y)
    = \sigma_2,\quad
    \sigma_{xy}(x, y)
    = 0.
\end{align}
\end{subequations}
This validates the correctness of the analytical method, as it recovers the expected uniform stress field for this trivial boundary condition.

\subsection{Stress distribution under uniaxial compression}\label{sec:uniaxial}

\subsubsection{Coefficient convergence and numerical stability}
We next analyze the case of uniaxial compression applied at the top and bottom edges:
\begin{equation}
    \sigma_1(y)=0,\quad
    \sigma_2(x)=-\sigma_\mathrm{ext}\delta(x).
    \label{eq:BC_uniaxial}
\end{equation}
The corresponding Fourier coefficients are:
\begin{equation}
    \widetilde{\sigma}_1^{(n)}
    = 0,\quad
    \widetilde{\sigma}_2^{(n)}
    = -\frac{\sigma_\mathrm{ext}}{L}.
    \label{eq:BC_uniaxial_Fourier}
\end{equation}
To solve the system, we truncate the series at $n = m = N_\mathrm{max}$ with $N_\mathrm{max}=128$, which provides sufficient numerical convergence. The transformed coefficients $\mathcal{A}_n$ and $\mathcal{C}_n$ [defined in Eq.~\eqref{eq:cal_A_C}] are computed from Eq.~\eqref{eq:X_solution}.

\begin{figure}[htbp]
    \centering
    \includegraphics[width=0.75\linewidth]{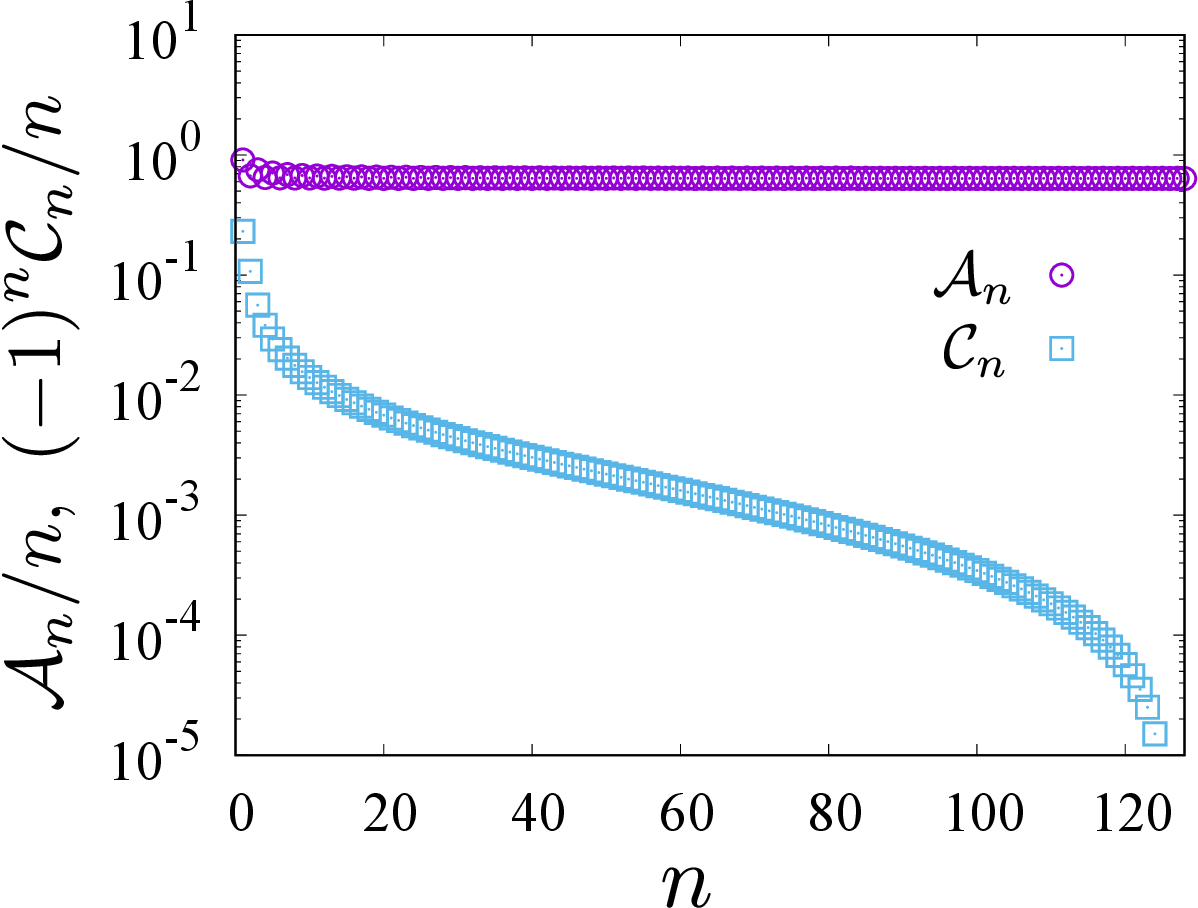}
    \vspace{1em}
    \caption{Plots of $\mathcal{A}_n$ and $\mathcal{C}_n$ against $n$ when we choose $N_\mathrm{max}=128$.}
    \label{fig:A_C}
\end{figure}

Figure \ref{fig:A_C} shows the plots of $\mathcal{A}_n$ and $\mathcal{C}_n$ determined from Eq.~\eqref{eq:X_solution} with Eq.~\eqref{eq:cal_A_C} when the boundary conditions are given by Eqs.~\eqref{eq:BC_uniaxial}. 
It is observed that $\mathcal{A}_n$ is almost constant and $\mathcal{C}_n$ converges to zero for $n\gg 1$. 
These behaviors show the convergence of the coefficients $A_n$ and $C_n$ for $n\gg1$ because both coefficients are related to each other via Eq.~\eqref{eq:cal_A_C}.

\begin{figure}[htbp]
    \centering
    \includegraphics[width=0.75\linewidth]{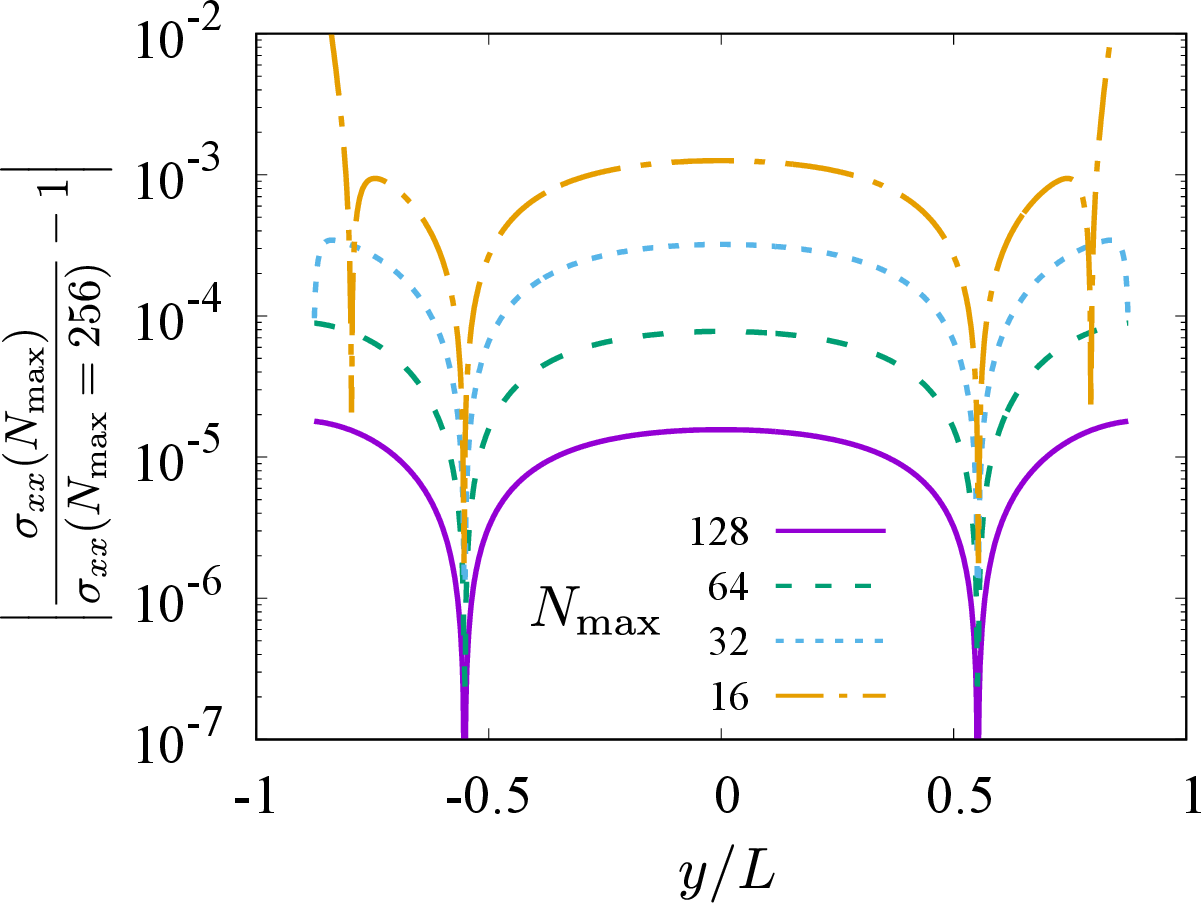}
    \vspace{1em}
    \caption{Convergence of the tensile stress component $\sigma_{xx}$ along the loading line $x=0$ for various $N_\mathrm{max}$. }
    \label{fig:convergence}
\end{figure}

This convergence behavior is further demonstrated in Fig.~\ref{fig:convergence}, where the spatial profiles of the tensile stress component $\sigma_{xx}$ along the loading line $x=0$ are plotted for different maximum truncation orders $N_\mathrm{max}=16$, $32$, $64$, $128$, and $256$. 
As $N_\mathrm{max}$ increases, the profiles become increasingly close to each other, and the results for $N_\mathrm{max}=128$ and $256$ overlap with the relative error $10^{-4}$. 
This indicates that the solution effectively converges by $N_\mathrm{max}=128$, justifying its choice for the subsequent analysis.

\begin{figure}[htbp]
    \centering
    \includegraphics[width=\linewidth]{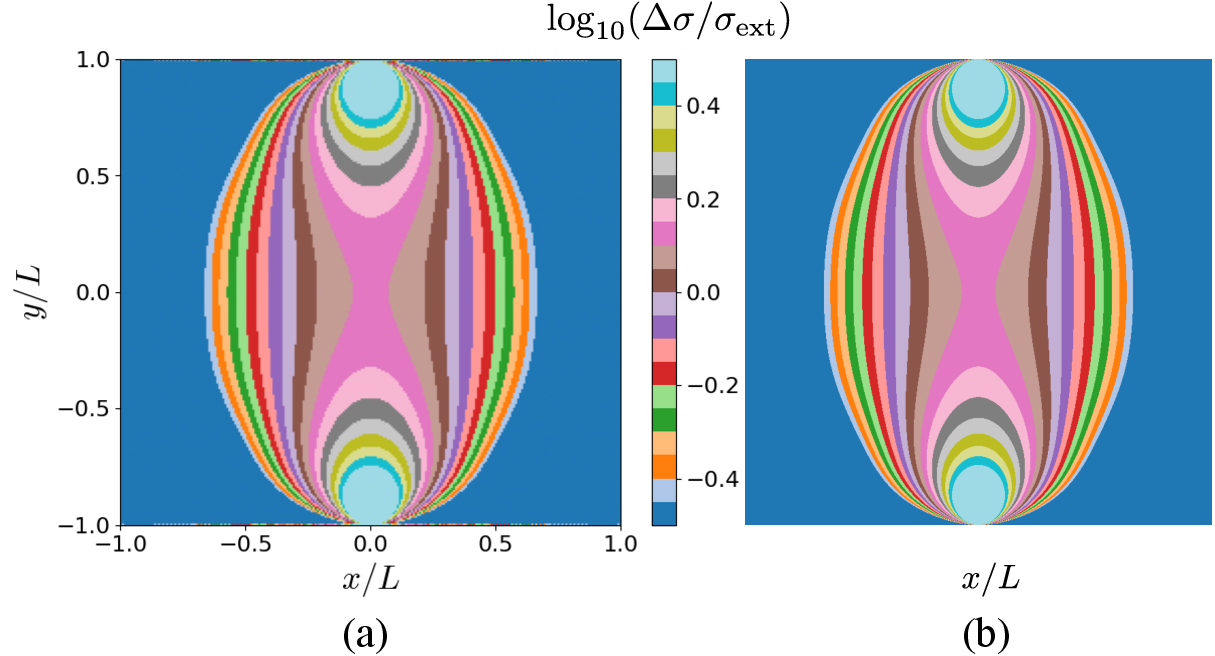}
    \caption{Plots of the principal stress difference $\Delta \sigma$ under the uniaxial compression obtained from (a) our theory and (b) the finite element analysis (see Appendix \ref{sec:FEM} for the detail).}
    \label{fig:result}
\end{figure}

Now, we plot the spatial profile of the stress. 
Figure \ref{fig:result} shows the principal stress difference
\begin{equation}
    \Delta \sigma 
    \equiv \sqrt{(\sigma_{xx} - \sigma_{yy})^2 + 4\sigma_{xy}^2}.
\end{equation}
See also Appendix \ref{sec:stress_component} for each component of the stress.
It should be noted that the quantity $\Delta \sigma$ is closely associated with the so-called fringe parameter in photoelasticity~\cite{Coker57}.
Specifically, the isocontours of the principal stress difference are known to align with the interference fringes observed in experimental photoelastic images.
The general profiles of the stress components closely resemble those obtained for circular domains~\cite{Timoshenko70, Coker57, Sato24_2D}, indicating a degree of universality in stress distribution under similar loading conditions.
Moreover, the present results show good agreement with experimental findings for square specimens~\cite{Zhao19, Manjit21, Saito52, Shimada54}, and the finite element analysis shown in Appendix \ref{sec:FEM}. 
These results additionally support the validity of the analytical formulation.

\subsubsection{Tensile stress profile on the loading line}
\begin{figure}[htbp]
    \centering
    \includegraphics[width=0.75\linewidth]{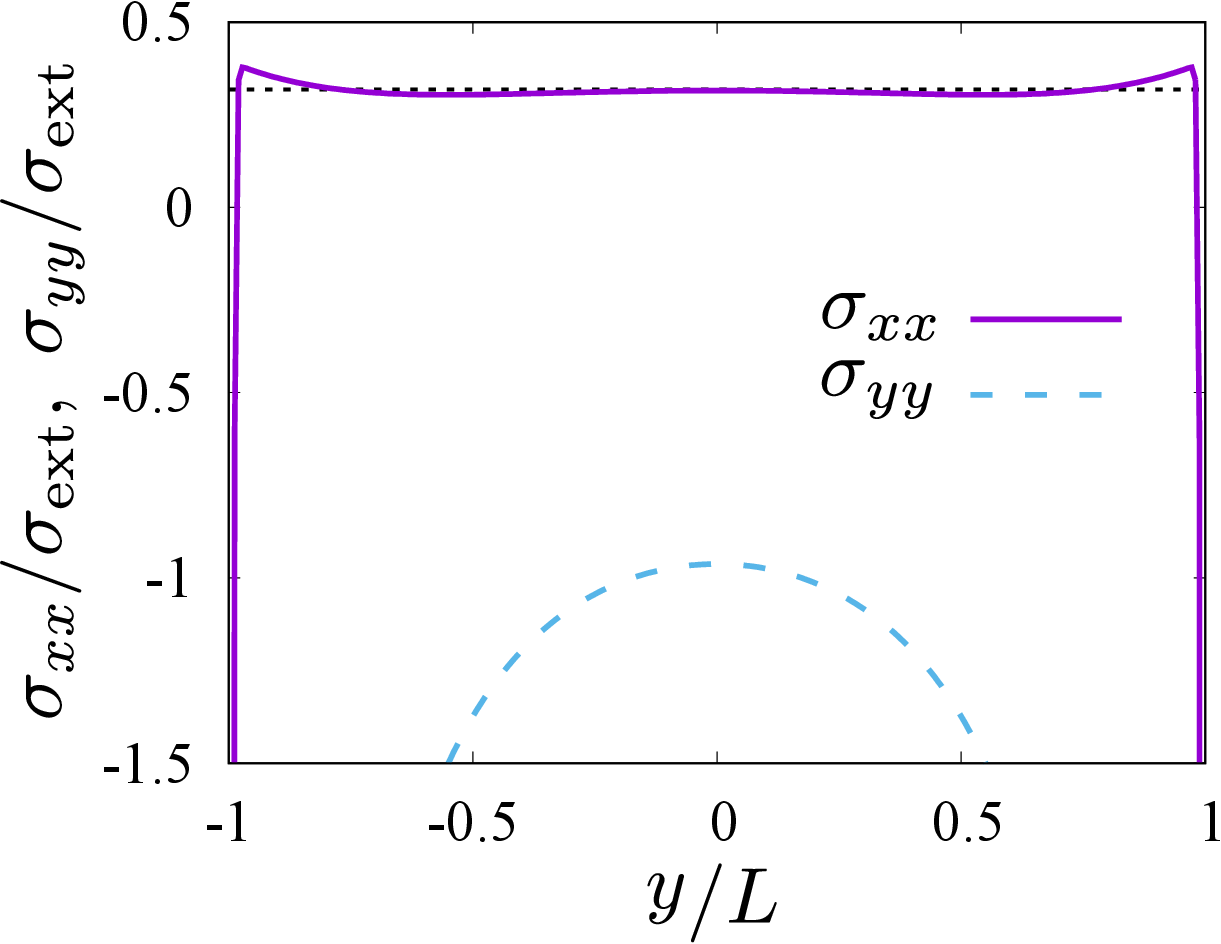}
    \caption{Plots of the tensile stress $\sigma_{xx}$ (solid line) and the compression stress $\sigma_{yy}$ (dashed line) on the loading line ($x=0$), where the dotted line represents $\sigma_{xx}/\sigma_\mathrm{ext}=1/\pi$ which is the tensile stress for a circular disk \eqref{eq:sigma_xx_circular}.}
    \label{fig:profile}
\end{figure}

To highlight the geometrical effects near the loading point, we plot $\sigma_{xx}$ and $\sigma_{yy}$ along the centerline $x=0$ in Fig.~\ref{fig:profile}. 
In the bulk region, $\sigma_{xx}$ remains constant, but it decreases near the load application point.

This deviation contrasts with the known solution for a circular disk, where the tensile stress is constant throughout:
\begin{equation}
    \frac{\sigma_{xx}}{\sigma_\mathrm{ext}} = \frac{1}{\pi}.
    \label{eq:sigma_xx_circular}
\end{equation}
In circular domains, the load is effectively screened near the contact region due to limited lateral coupling. 
In contrast, for a square cross-section, the surrounding material provides sufficient resistance, leading to localized compression around the load. 
This causes a non-uniform $\sigma_{xx}$ near the top edge, capturing a realistic feature not evident in circular solutions.

\subsection{Stress distribution under biaxial compression}\label{sec:biaxial}
\subsubsection{Combined effect of orthogonal concentrated loads}
We now consider the case where concentrated compressive forces are applied in both directions:
\begin{equation}
    \sigma_1(y) = -\sigma_{1,\mathrm{ext}}\delta(y),\quad
    \sigma_2(x) = -\sigma_{2,\mathrm{ext}}\delta(x).
\end{equation}
The Fourier coefficients are thus:
\begin{equation}
    \widetilde{\sigma}_1^{(n)}
    = -\frac{\sigma_{1,\mathrm{ext}}}{L},\quad
    \widetilde{\sigma}_2^{(n)}
    = -\frac{\sigma_{2,\mathrm{ext}}}{L}.
\end{equation}
The resulting coefficients $\mathcal{A}_n$ and $\mathcal{C}_n$ are again computed from Eq.~\eqref{eq:X_solution} with the same truncation $N_\mathrm{max}=128$.

\subsubsection{Stress field patterns and anisotropic amplification}

\begin{figure}[htbp]
    \centering
    \includegraphics[width=0.75\linewidth]{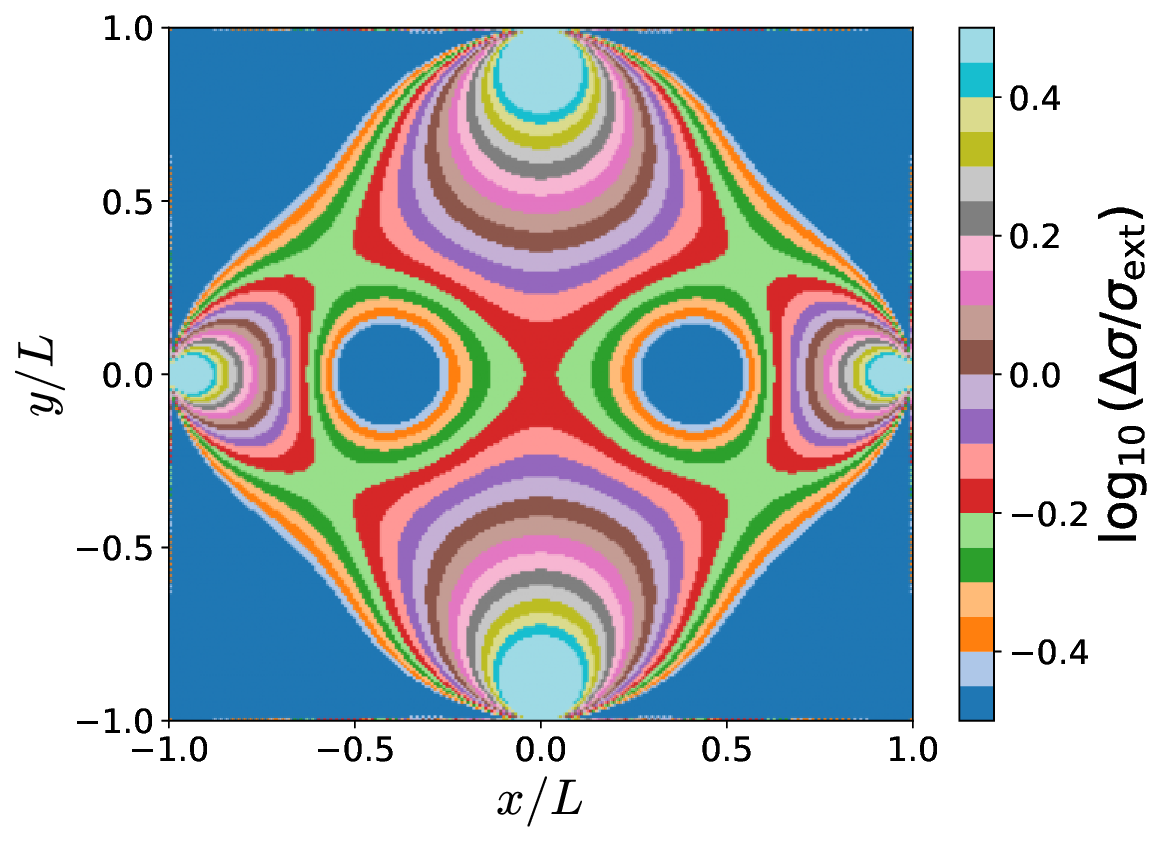}
    \caption{Plot of the principal stress difference $\Delta \sigma$ under the biaxial compression.}
    \label{fig:result_biaixial}
\end{figure}

Figure~\ref{fig:result_biaixial} shows the stress distribution for the case $\sigma_{1,\mathrm{ext}} / \sigma_{2,\mathrm{ext}} = 1/2$.
See also Appendix \ref{sec:stress_component} for each component of the stress.
The obtained stress field reflects a superposition of the two uniaxial compression patterns. 
In particular, the principal stress difference $\Delta \sigma$ reveals amplification along the $y$-axis and attenuation along the $x$-axis due to the asymmetric loading.
This behavior illustrates the anisotropic stress propagation under biaxial concentrated loads and further emphasizes the interplay between geometry and boundary conditions.

\subsection{Discussion}
The three loading scenarios presented above demonstrate how the stress field varies depending on the nature and symmetry of the applied loads.
(i) In the case of uniform biaxial tension, the stress distribution is trivial and uniform, serving as a benchmark to validate the method.
(ii) Under uniaxial concentrated compression, the stress field exhibits strong localization near the loading edge and significant deviation from the circular case, revealing the effect of square geometry on stress transmission.
(iii) For biaxial compression, the solution behaves as a linear superposition of uniaxial solutions, but the resulting stress patterns reveal directional amplification and anisotropy not present in simpler geometries.

These findings emphasize the importance of considering both boundary conditions and specimen geometry when analyzing internal stress fields. 
The presented method captures these features effectively and may serve as a useful basis for further analytical or numerical studies in elasticity and fracture mechanics.

In addition, similar considerations arise in practical engineering problems such as the influence of shear strain on girder deflection~\cite{Lazarevic24}, the thermal response of steel structures under fire loading~\cite{Major24}, and the stiffness and damping behavior of steel dampers subjected to lateral cyclic loading~\cite{Ampangallo24}.
Our analytical framework could potentially be extended to these contexts, providing theoretical benchmarks and deeper insight into the interplay between geometry, loading conditions, and material response.

\section{Conclusion}\label{sec:conclusion}
In this study, we have developed an analytical framework to evaluate the internal stress distributions in an elastic body with a square cross-section under biaxial tension or compression.
By employing the Airy stress function formalism and satisfying the biharmonic equation, we constructed stress fields that rigorously meet the prescribed boundary conditions.

The boundary value problem was solved by determining a set of Fourier coefficients that enforce the external loading.
This allowed us to obtain closed-form stress distributions under various loading conditions.
In particular, we examined uniaxial and biaxial concentrated compressive loads as illustrative examples.
For the uniaxial case, the calculated stress fields exhibited strong agreement with known photoelastic fringe patterns and prior theoretical results, confirming the validity of the formulation.
Under biaxial loading, the superposition of orthogonal uniaxial stress components led to a rich spatial variation in the principal stress difference, demonstrating how directional loading asymmetry manifests within the internal stress field.

The analytical approach introduced here offers a versatile platform for exploring stress distributions in other geometries.
One natural extension is to rectangular domains with arbitrary aspect ratios, where the differing spatial frequencies in the $x$- and $y$-directions require modified solution techniques.
Although the method was derived here for the square case, its mathematical structure can be adapted to rectangles, with Saint-Venant’s principle suggesting that the bulk stress distribution away from boundaries would remain accurately captured.
Another promising direction is the analysis of elliptical or irregularly shaped domains, which are relevant to both fundamental elasticity theory and applied problems in structural and materials engineering.

\backmatter

\bmhead{Acknowledgements}
This work is partially supported by the Grant-in-Aid of MEXT for Scientific Research (Grant No.~JP24K06974, No.~JP24K07193, and No.~JP25K01063).

\bmhead{Conflict of interest}
The authors declare no competing interests.

\appendix
\section{General solution of the biharmonic equation~(\ref{eq:biharmonic})}\label{sec:biharmonic}
In this Appendix, let us show the general solution of the biharmonic equation~\cite{Timoshenko70}.
First, let us consider the solution of the harmonic equation:
\begin{equation}
    \nabla^2 \phi = 0.
    \label{eq:harmonic}
\end{equation}
Assuming the separation of variables as $\phi(x, y)=X(x)Y(y)$, Eq.~\eqref{eq:harmonic} becomes
\begin{equation}
    \frac{\mathrm{d}^2 X(x)}{\mathrm{d}x^2}
    = -\frac{\mathrm{d}^2 Y(y)}{\mathrm{d}y^2}.
    \label{eq:X_Y_eq}
\end{equation}
Because the left-hand and right-hand sides of Eq.~\eqref{eq:X_Y_eq} depend only on $x$ and $y$, respectively, Eq.~\eqref{eq:X_Y_eq} becomes constant~\cite{Timoshenko70}.
Then, the general solution becomes~\cite{Timoshenko70}
\begin{align}
    \phi(x, y)
    &= \int_{-\infty}^\infty \mathrm{d}k 
    \left[
    \mathcal{A}_\mathrm{cc}(k)\cos(kx)\cosh(ky)
    + \mathcal{A}_\mathrm{cs}(k)\cos(kx)\sinh(ky)\right.\nonumber\\
    &\left.\hspace{5em}
    + \mathcal{A}_\mathrm{sc}(k)\sin(kx)\cosh(ky)
    + \mathcal{A}_\mathrm{ss}(k)\sin(kx)\sinh(ky)\right.\nonumber\\
    &\left.\hspace{5em}
    + \mathcal{B}_\mathrm{cc}(k)\cosh(kx)\cos(ky)
    + \mathcal{B}_\mathrm{cs}(k)\cosh(kx)\sin(ky)\right.\nonumber\\
    &\left.\hspace{5em}
    + \mathcal{B}_\mathrm{sc}(k)\sinh(kx)\cos(ky)
    + \mathcal{B}_\mathrm{ss}(k)\sinh(kx)\sin(ky)\right],
    \label{eq:solution_harmonic}
\end{align}
where $\mathcal{A}_\mathrm{cc}(k)$, $\mathcal{A}_\mathrm{cs}(k)$, $\mathcal{A}_\mathrm{sc}(k)$, $\mathcal{A}_\mathrm{ss}(k)$, $\mathcal{B}_\mathrm{cc}(k)$, $\mathcal{B}_\mathrm{cs}(k)$, $\mathcal{B}_\mathrm{sc}(k)$, and $\mathcal{B}_\mathrm{ss}(k)$ are the coefficients to be determined by the boundary conditions.

Once we obtain the solution of the harmonic equation $\phi(x, y)$ as Eq.~\eqref{eq:solution_harmonic}, it is known that 
\begin{equation}
    x\phi(x, y),\ y\phi(x, y),
\end{equation}
satisfy the biharmonic equation \eqref{eq:biharmonic}.
In addition, it is easy to see that $x^2$ and $y^2$ are also solutions to Eq.~\eqref{eq:biharmonic}.

Then, the stress components can be calculated via Eq.~\eqref{eq:sigma_phi}.
Here, we are only interested in the case where the system is symmetric with respect to the $x$- and $y$-axes.
In addition, $\sigma_{xy}$ should vanish at $x=\pm L$ and at $y=\pm L$.
This means that the wave number $k$ becomes discrete and only the form $n\pi/L$ (where $n$ is an integer) is allowed.
Therefore, we obtain the expression of Eq.~\eqref{eq:Airy}.

\section{Each component of the stress under uniaxial and biaxial cases}\label{sec:stress_component}
In the main text, we have only shown the results of the principal stress difference $\Delta \sigma$.
In this Appendix, we present the plot of the each component of the stress obtained from our theory.

\begin{figure}[htbp]
    \centering
    \includegraphics[width=0.49\linewidth]{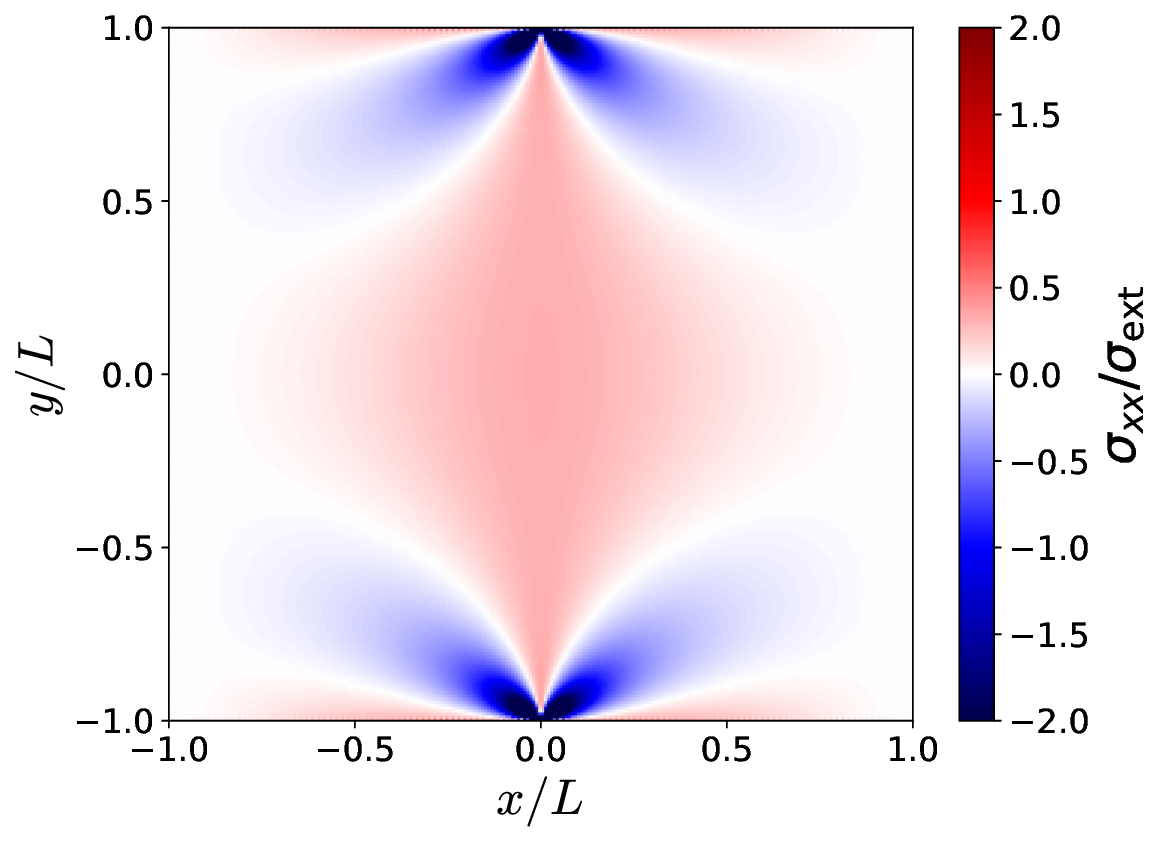}
    \includegraphics[width=0.49\linewidth]{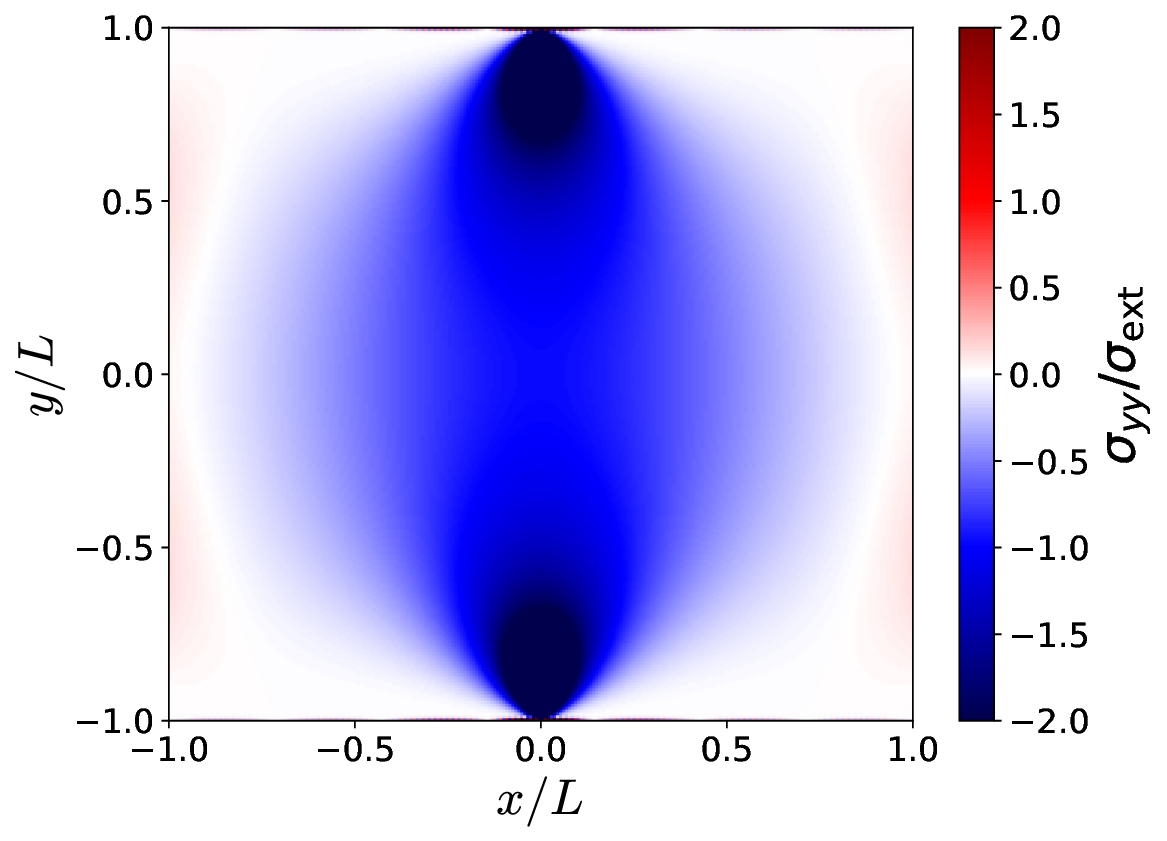}
    \includegraphics[width=0.49\linewidth]{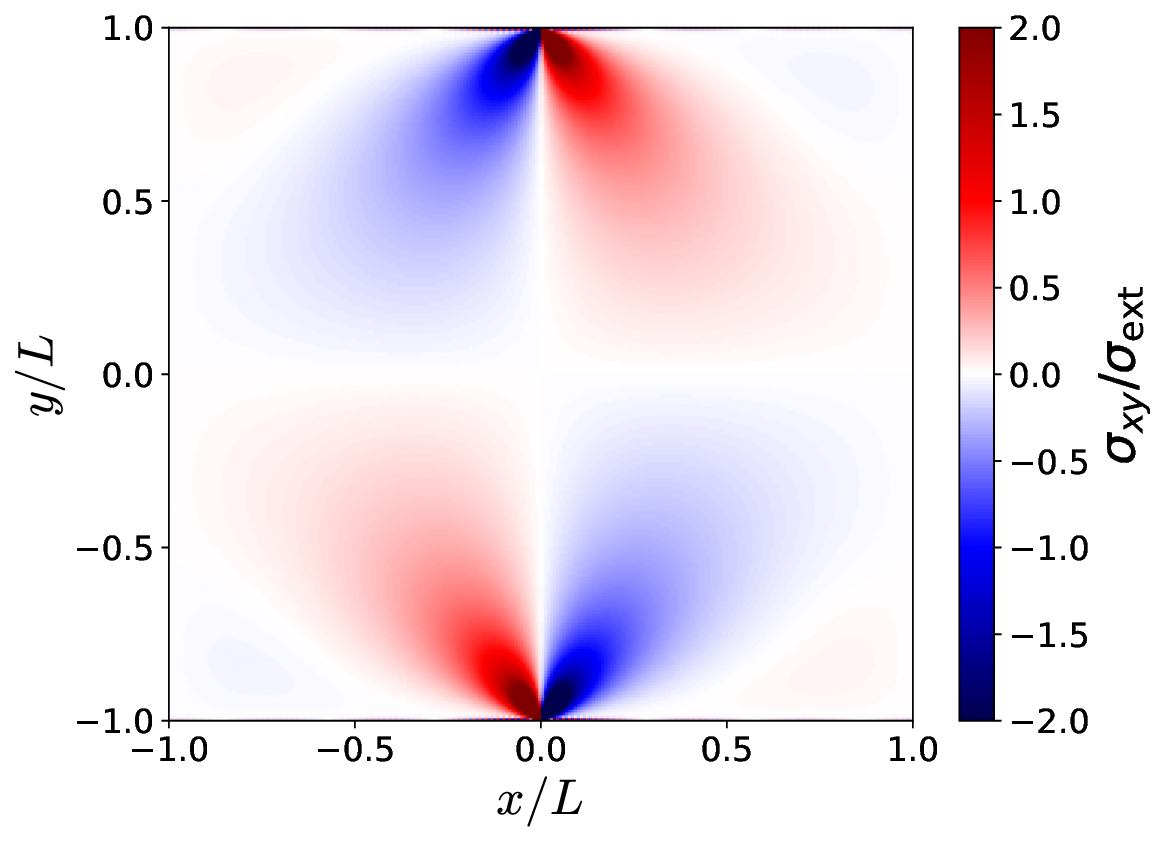}
    \caption{Plots of (top left) $\sigma_{xx}$, (top right) $\sigma_{yy}$, and (bottom) $\sigma_{xy}$ under the uniaxial compression.}
    \label{fig:result_uni_component}
\end{figure}

\begin{figure}[htbp]
    \centering
    \includegraphics[width=0.49\linewidth]{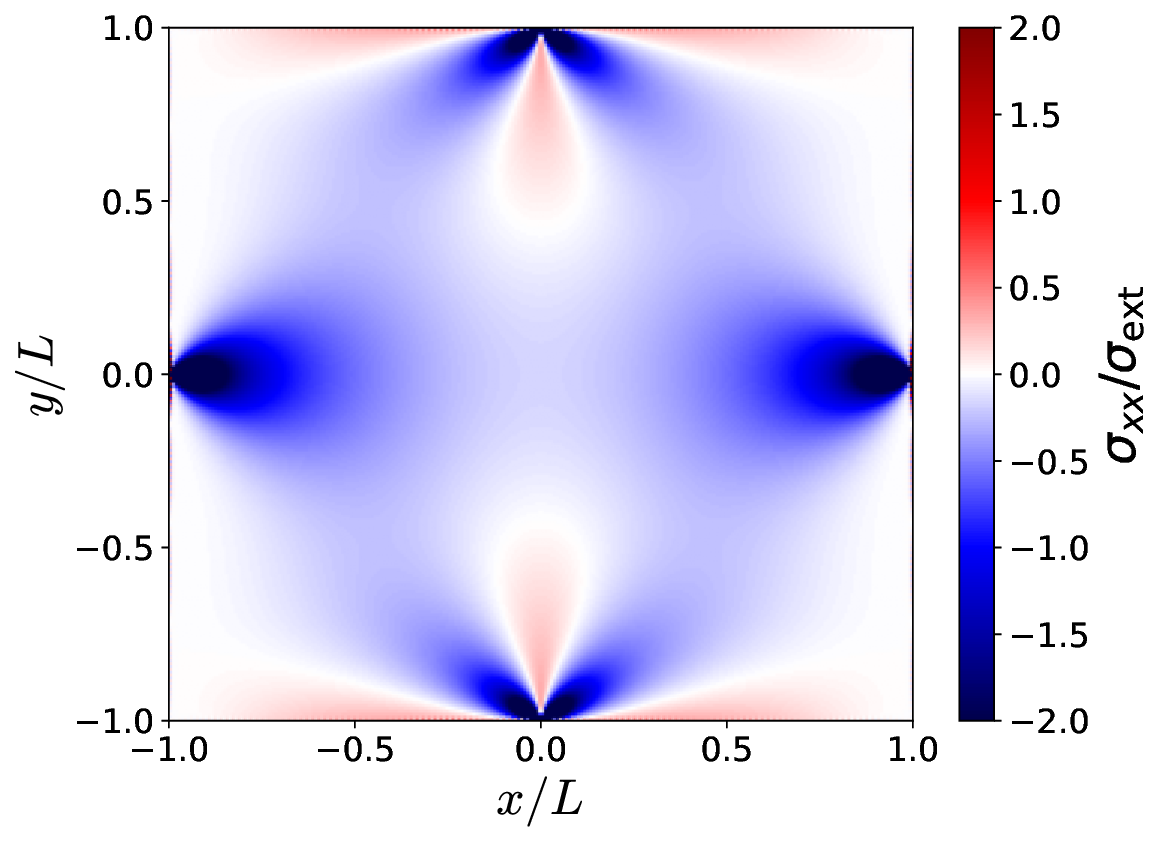}
    \includegraphics[width=0.49\linewidth]{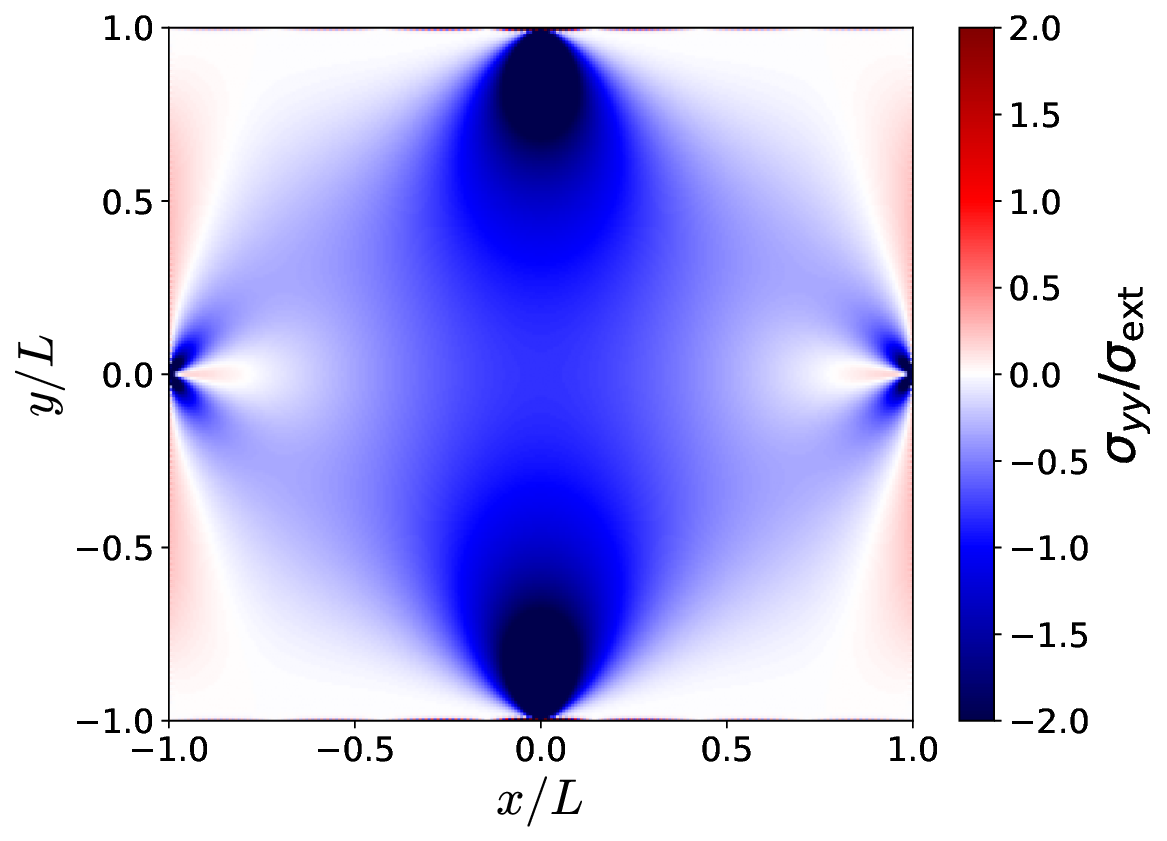}
    \includegraphics[width=0.49\linewidth]{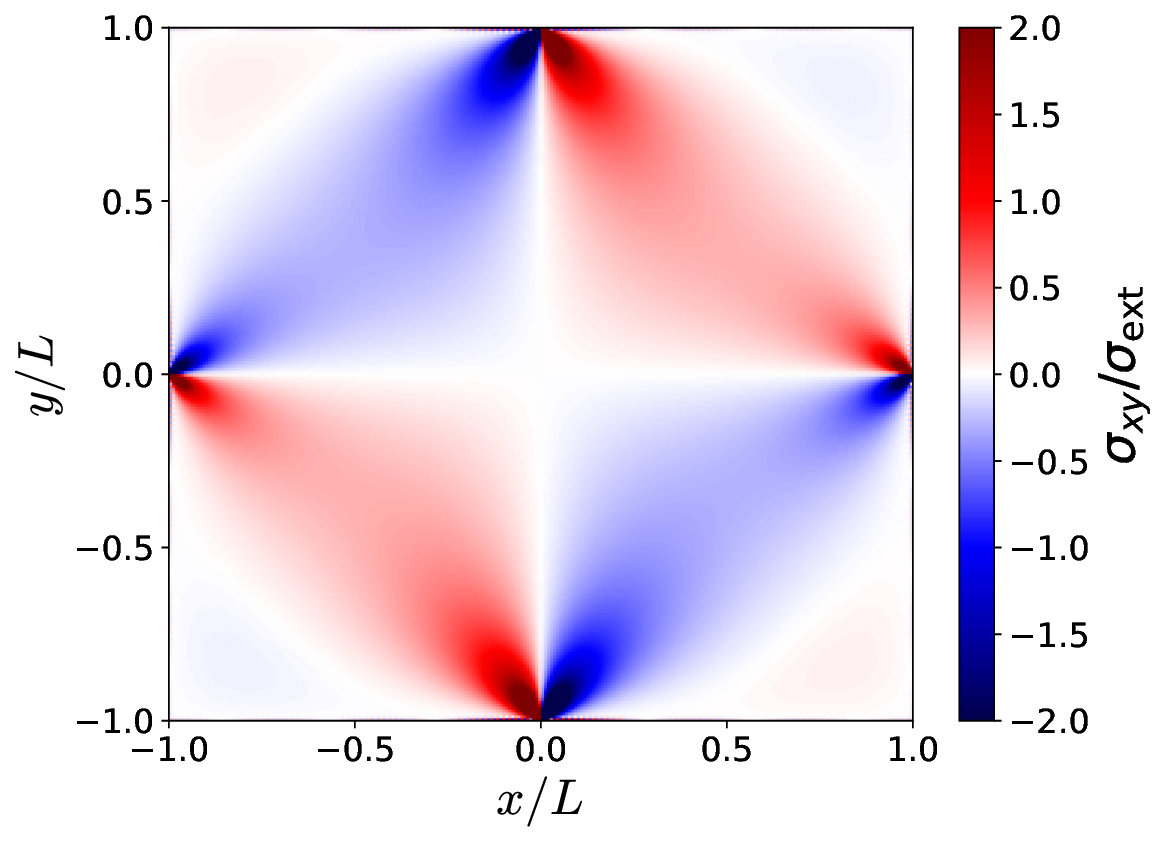}
    \caption{Plots of (top left) $\sigma_{xx}$, (top right) $\sigma_{yy}$, and (bottom) $\sigma_{xy}$ under the biaxial compression.}
    \label{fig:result_bi_component}
\end{figure}

Figures \ref{fig:result_uni_component} and \ref{fig:result_bi_component} show the plots of the stress components under the uniaxial and biaxial compression, respectively.
In both cases, $\sigma_{xy}$ becomes zero at $x=\pm L$ and $y=\pm L$, which satisfy the boundary condition \eqref{eq:BC}.

\section{Validity of the present theory by comparison with the finite element method}\label{sec:FEM}
To verify the validity of the present theoretical formulation, we consider a special case in which the square cross section is subjected to uniaxial tension along the $y$-direction as shown in Sect.~\ref{sec:uniaxial}. 
In this case, the analytical solution derived from our theory is evaluated and compared with numerical results obtained from the finite element method (FEM).

The FEM analysis is performed using MATLAB's built-in finite element solver. 
The computational mesh is automatically generated by the software, and no manual refinement is applied. 
Boundary conditions are the same as Eq.~\eqref{eq:BC_uniaxial}.
Symmetry conditions are imposed along the remaining faces to prevent rigid-body motion. 
Other material properties in the FEM model are also identical to those assumed in the analytical formulation.

Figure~\ref{fig:result}(b) shows the distribution of the pricipal stress difference $\Delta\sigma$ obtained from the present theory and from the FEM analysis. 
The results exhibit excellent agreement across the entire cross section
This close correspondence confirms that the present theory accurately captures the stress distribution in the uniaxial loading case, thereby supporting the validity of the theoretical framework developed in this study.

\bibliography{References}

\end{document}